\documentclass[11pt]{article}
\usepackage{amsmath,amsfonts,amssymb,amsthm}
\usepackage[margin=1in]{geometry}
\usepackage[colorlinks]{hyperref}
\usepackage{graphicx}

\begin{document}


\title{Operational criterion for controlled dense coding with non-trivial tri-partite entangled states}

\author{Sovik Roy$^{1}$
\thanks{sovikr@rediffmail.com} \\
$^1$ Department of Mathematics, EM 4/1, Techno India, Salt Lake, Kolkata - 91, India\\\\
Biplab Ghosh$^{2}$ 
\thanks{biplab1976@gmail.com}\\
$^2$ Department of Physics, Vivekananda College for Women, Barisha, Kolkata - 8, India\\\\
Md. Manirul Ali$^{3}$
\thanks{maniquantum@gmail.com}\\
$^{3}$ Physics Division, National Center for Theoretical Sciences, National Tsing Hua University,\\ Hsinchu 30013, Taiwan
}

\date{\today}
\maketitle

\begin{abstract}
\noindent In this paper, we provide an operational criterion for controlled dense coding with a general class of three-qubit partially entangled states. A general three-qubit pure entangled state can be classified into two inequivalent classes according to their genuine tripartite entanglement. We claim that if a three-qubit state shows entanglement characteristic similar to $GHZ$-class then such non-trivial tripartite states are useful in {\it controlled dense coding} whereas states belonging to $W$-class are not useful for that. We start with a particular class of non-trivial partially entangled states belonging to $GHZ$-class and show that they are effective in controlled dense coding. Then we cite several other examples of different types of tripartite entangled states to support our conjecture. 
\end{abstract}


\section{Introduction}
Ever since its inception, the study of entanglement has always been in lime-light.  Entanglement plays a pivotal role in quantum information
processing tasks. Among all such tasks, dense coding \cite{bennett691992} is one such application of entanglement, first implemented experimentally
with polarized entangled photons \cite{mattle761996}. The fundamental idea of dense coding is that, with the help of shared entangled channel
between the two parties, two bits of information can be transmitted from sender to the receiver using only one qubit, provided the shared channel
is maximally entangled. For the sake of security purpose, this transmission of bits from sender to the receiver can however be controlled by another
third party when the trio share some tripartite entangled state among them. Such a modified dense coding protocol is known as controlled dense
coding (CDC) scheme. The scheme was first presented by using the three-qubit maximally entangled Greenberger-Horne-Zeilinger (GHZ) state
in \cite{haoli054301}.  $GHZ$-state can be generated  in the laboratory and have been demonstrated experimentally using two pairs of entangled photons and 
also one can make optimal distillation of these states \cite{boum1345,acin4811}. Another important tripartite entangled state is $W$-state \cite{durr622000}.  In real experimental situation it has always been a challenge to obtain such multi-qubit maximally entangled resources
\cite{bennett531996}. Recently, X-P. Zang \textit{et.al} have shown a fantastic way of transforming a bipartite non-maximally entangled states in to a tripartite $W$-state in cavity QED \cite{xpz862016}. Also a scheme for deterministic joint remote preparation of an arbitrary four qubit $W$-type entangled state has been discussed by H. Fu \textit{et. al} in \cite{hfu882017}. Hence, it is of fundamental importance to identify the multi-qubit partially entangled systems which may be useful in tasks
like CDC. Controlled dense coding has been investigated with four-qubit entangled states \cite{fuxializh}, three-qubit symmetric and non-symmetric states \cite{you482297,liu2015}, three qubit maximal sliced states \cite{liumo2182}, and with six-particle graph state \cite{Dliusix}. Controlled dense coding has also been realized using a
five atom cluster state in cavity-QED \cite{you1851}. However, it was shown in \cite{roy1550033} that the $W$-class of states are not useful 
in controlled dense coding. Since $GHZ$- and $W$- states are two in-equivalent classes of tripartite maximally entangled states, our main motivation in this work is to put forward a 
conjecture where we claim that, the states belonging to $GHZ$- class will be useful in CDC whereas those belonging to $W$- class will be not. This is the main impetus of this work which will be having incredible experimental relevance. For the sake of further discussion, in this paper, we shall use the term $`$ trivial' for $GHZ$- and $W$- states and $`$non-trivial' for the tripartite states other than $GHZ$- states and $W$- states.\\

\noindent We therefore classify the tripartite states into two inequivalent classes according to their genuine tripartite entanglement. A$  $n appealing feature of multipartite entanglement in quantum information science is the existence of two different kinds of genuine tripartite entanglement, viz. entanglement of $GHZ$-class and entanglement of $W$-class of states \cite{ghzw1,ghzw2}. Entanglement properties of these two classes of states are very interesting. While $GHZ$-class of states are fragile in nature with respect to loss of particles, the entanglement of $W$-class of states is maximally robust under loss of particles. The in-equivalence of these two class of states under stochastic local operation and classical communication (SLOCC) has also been extensively discussed in \cite{durr622000}. This in-equivalence implies that if two arbitrary states
$\vert \Phi\rangle$ and $\vert \Psi\rangle$ can respectively be converted into states belonging
to $GHZ$-class or $W$-class then it will not be possible to transform $\vert \Phi\rangle$ into $\vert \Psi\rangle$ or vice versa. The entanglement
of $GHZ$-class and $W$-class of states are contrasting in nature. We, with some examples, in this paper show that if a tripartite state shows entanglement
characteristic similar to $GHZ$-class then those states are useful in controlled dense coding under suitable choice of the state parameters. Our
main motivation in this work is therefore to classify the non-trivial classes of tripartite entangled states which are valuable in CDC. The focus in this work is to categorize the class of three-qubit non-trivial partially entangled states into either of the two genuine tripartite class of states ($GHZ$ and $W$) on the basis of their usefulness in CDC. This will provide our (conjectured) operational criteria based on which one can differentiate the non-trivial classes of tripartite entangled states that are useful for CDC from those
that are not useful in CDC. The paper is thereby organized as follows. In Sec~\ref{sec:XiStates}, we begin our analysis of controlled dense coding with a
class of general tripartite partially entangled states which were defined in \cite{li052305} and consequently study their effectiveness in CDC. Section~\ref{sec:TangleXi} discusses the entanglement properties of the class of partially entangled states of reference \cite{li052305}. We introduce our (conjectured) operational criteria for states that are useful for CDC in Sec~\ref{sec:Criteria}  To validate our claim, we cite several
other examples of non-trivial tripartite entangled states which are useful in CDC in Sec~\ref{sec:example} Finally, a conclusion is given in Sec~\ref{sec:conclusion}  
\section{Class of partially entangled states in controlled dense coding}\label{sec:XiStates}
A class of partially entangled superposition of Bell states was constructed and was shown to be useful for perfect
controlled teleportation by Xi \textit{et.al} \cite{li052305}. The class of states are defined as
\begin{eqnarray}
\label{xi1}
\vert \chi\rangle_{abc} = \alpha ~\vert 0\rangle_{a} ~\vert \phi^{+}\rangle_{bc} + \beta ~\vert 1\rangle_{a} ~\sigma_{kc}
~\vert \phi^{+}\rangle_{bc}, \\
\label{xi2}
\vert {\widetilde \chi} \rangle_{abc} = \alpha ~\vert 0\rangle_{a} ~\vert \phi^{-}\rangle_{bc} + \beta ~\vert 1\rangle_{a} ~\sigma_{kc}
~\vert \phi^{-}\rangle_{bc},
\end{eqnarray}
where $|\alpha|^{2} + |\beta|^{2} = 1$ and $\vert \phi^{\pm}\rangle = \left( \vert 00\rangle \pm \vert 11\rangle \right)/\sqrt{2}$.
The qubits $a$, $b$, $c$ are distributed to Alice, Bob and Cliff respectively and $\sigma_{kc}$, ($k = x,y,z$) are respectively three
Pauli spin matrices acting on the qubit $c$, where
\begin{eqnarray}
\sigma_{xc} = \left(\begin{array}{cc}
 0  & 1\\
1 & 0\\
\end{array}
\right),
i\sigma_{yc} = \left(\begin{array}{cc}
 0  & -1\\
1 & 0\\
\end{array}
\right),
\sigma_{zc} = \left(\begin{array}{cc}
 1  & 0\\
0 & -1\\
\end{array}
\right).
\label{p_3}
\end{eqnarray}
In this paper we ask the question: ``whether the states (\ref{xi1}) and (\ref{xi2}) are useful for controlled dense coding too !''
In controlled dense coding scheme, when the trio share some entangled channels among them, then depending upon any one party's
von-Neumann measurement outcomes, the remaining two parties are expected to share maximal Bell states so that usual dense
coding can be pursued. That ``one party'' acts as the controller of the scheme and the controller can be any one of the trios. The
remaining two parties play the roles of sender and receiver of the bits of information. To illustrate the scheme, we start with
the state (\ref{xi1}), and assume Alice, who possesses qubit $a$ as the controller and Bob and Cliff possess respectively the
qubits $b$ and $c$. Now if Alice decides to measure with respect to measurement basis
$\{\vert 0\rangle_{a}, \vert 1\rangle_{a}\}$, then if her von-Neumann measurement results in $\vert 0\rangle_{a}$, Bob and
Cliff know that they would share the maximal Bell state $\vert \phi^{+}\rangle$, of course when Alice sends her measurement
outcomes to both Bob and Cliff. On the other hand, if she gets her von-Neumann outcome as $\vert 1\rangle_{a}$, then Bob and
Cliff would share any one of the maximally entangled Bell states for an appropriate choice of Pauli spin operators.
Thus, depending upon Alice's measurement result $\vert 1\rangle_{a}$, Bob and Cliff know which Bell state they are going to share
\begin{eqnarray}
\sigma_{kc}\vert \phi^{+}\rangle_{bc}|_{k\:=\: x} \rightarrow \vert \psi^{+}\rangle_{bc}, \\
\sigma_{kc}\vert \phi^{+}\rangle_{bc}|_{k\:=\: y} \rightarrow \vert \psi^{-}\rangle_{bc}, \\
\sigma_{kc}\vert \phi^{+}\rangle_{bc}|_{k\:=\: z} \rightarrow \vert \phi^{-}\rangle_{bc},
\end{eqnarray}
where $\vert \psi^{\pm}\rangle = \left( \vert 01\rangle \pm \vert 10\rangle \right)/\sqrt{2}$. So we see that, in all these cases
Bob and Cliff will be sharing any one of the four Bell states between them and hence they can proceed with the usual dense coding
scheme. Just as in equations (\ref{xi1}) and (\ref{xi2}), one can also start with a similar class of partially entangled
states as shown below and can proceed with the protocol mentioned above with these class of states.
\begin{eqnarray}
\label{xi3}
\vert \xi \rangle_{abc} = \alpha ~\vert 0\rangle_{a} ~\vert \psi^{+}\rangle_{bc} + \beta ~\vert 1\rangle_{a} ~\sigma_{kc}
~\vert \psi^{+}\rangle_{bc}, \\
\label{xi4}
\vert {\widetilde \xi} \rangle_{abc} = \alpha ~\vert 0\rangle_{a} ~\vert \psi^{-}\rangle_{bc} + \beta ~\vert 1\rangle_{a} ~\sigma_{kc}
~\vert \psi^{-}\rangle_{bc}.
\end{eqnarray}


\section{Entanglement of the class of partially entangled states}\label{sec:TangleXi}

In order to analyze the entanglement characteristics of the partially entangled states (\ref{xi1}),  (\ref{xi2}), (\ref{xi3}) and (\ref{xi4}),
we use the 3-tangle $(\tau)$ as a measure for genuine tripartite entanglement \cite{coffman612000}. The 3-tangle $(\tau)$ is
defined as
\begin{eqnarray}
\tau = C^{2}_{a(bc)} - C^{2}_{ab} - C^{2}_{ac}
\label{tangle},
\end{eqnarray}
where $C^{2}_{a(bc)}$ measures the entanglement between qubit $a$ and the joint state of qubits $b$ and $c$ and $C^{2}_{ab}$ (or $C^{2}_{ac}$) measures the entanglement between qubits $a$ and $b$ (or between qubits $a$ and $c$). $C^{2}$ here means \textit{concurrence squared}. For pure entangled state one can show that $C_{a(bc)} = 2\:\sqrt{\det \rho_{a}}$ while $\rho_{abc}$ is the corresponding three-qubit density matrix. The tripartite entangled state (\ref{xi1}) can be explicitly expressed with the choices of state parameters  $\alpha=\sin\:\epsilon$ and $\beta=\cos\:\epsilon$ as follows
\begin{eqnarray}
\vert \chi^{1}\rangle_{abc} =  \frac{\sin\epsilon}{\sqrt{2}}(\vert 000\rangle_{abc} + \vert 011\rangle_{abc}) +\nonumber\\
\frac{\cos\epsilon}{\sqrt{2}}(\vert 110\rangle_{abc} + \vert 101\rangle_{abc}),
\label{xi1a}
\end{eqnarray}
\begin{eqnarray}
\vert \chi^{2}\rangle_{abc} =  \frac{\sin\epsilon}{\sqrt{2}}(\vert 000\rangle_{abc} + \vert 011\rangle_{abc}) +\nonumber\\
\frac{\cos\epsilon}{\sqrt{2}}(\vert 101\rangle_{abc} - \vert 110\rangle_{abc}),
\label{xi1b}
\end{eqnarray}
\begin{eqnarray}
\vert \chi^{3}\rangle_{abc} =  \frac{\sin\epsilon}{\sqrt{2}}(\vert 000\rangle_{abc} + \vert 011\rangle_{abc}) +\nonumber\\
\frac{\cos\epsilon}{\sqrt{2}}(\vert 100\rangle_{abc} - \vert 111\rangle_{abc}).
\label{xi1c}
\end{eqnarray}
The partially entangled states of equations (\ref{xi1a}), (\ref{xi1b}) and (\ref{xi1c}) are associated with different
choices of the Pauli spin operators $\sigma_{xc}$, $\sigma_{yc}$ and $\sigma_{zc}$ respectively in equation (\ref{xi1}).
The genuine tripartite entanglement or 3-tangle (\ref{tangle}) for each of these partially entangled states is found to be
$\sin^{2}(2\epsilon)$. The states (\ref{xi2}), (\ref{xi3}) and (\ref{xi4}) can also be shown to be useful in CDC and
having 3-tangle of the same form as that of state (\ref{xi1}).

\section{Operational criterion for controlled dense coding}\label{sec:Criteria}

Given a pure nontrivial tripartite entangled state, our aim is to find an operational criterion for this state to be used in controlled dense coding. Now it is known that
for a  general nontrivial tripartite state (whose density matrix is denoted as $\rho_{abc}$), if the local ranks of all the reduced
density matrices is $1$, then the state $\rho_{abc}$ belongs to product-class state ($a$-$b$-$c$) whereas if the ranks of any
two reduced system is $2$ and the other one has rank $1$, then the tripartite state falls into the class of bi-separable states
($a$-$bc$ or $b$-$ca$ or $c$-$ab$). The interesting scenario is however when $rank(\rho_{a})=rank(\rho_{b})
= rank(\rho_{c}) = 2$, and there are two in-equivalent classes of tripartite entangled states which satisfy this condition
\cite{durr622000}. It was shown  that any non-trivial tripartite pure entangled state can be characterized into these
two inequivalent classes known as $GHZ$-class and $W$-class of states \cite{durr622000}. The most general form of
states belonging to the in-equivalent $GHZ$-class is given by
\begin{eqnarray}
\label{ghzClass}
\vert \psi_{GHZ} \rangle_{abc} = \sqrt{K}(\cos\:\delta\:|0\rangle_{abc}|0\rangle_{abc}|0\rangle_{abc}+\nonumber\\ \sin\:\delta\:e^{i\:\varphi}\:|\varphi\rangle_{a}|\varphi\rangle_{b}|\varphi\rangle_{c}),\\
\nonumber
&&{}
\end{eqnarray}
where,
\begin{eqnarray}
\label{extra}
|\varphi\rangle_{a} &=& \cos\:\alpha\:|0\rangle + \sin\:\alpha\:|1\rangle,\nonumber{}\\
|\varphi\rangle_{b} &=& \cos\:\beta\:|0\rangle + \sin\:\beta\:|1\rangle,\nonumber{}\\
|\varphi\rangle_{c} &=& \cos\:\gamma\:|0\rangle + \sin\:\gamma\:|1\rangle,\nonumber{}\\
\end{eqnarray}
and  $K=(1+2\cos\delta\sin\delta\cos\alpha\cos\beta\cos\gamma\cos\varphi)^{-1}\in \left(\frac{1}{2},\infty \right)$ is a normalization factor. The ranges of the five parameters of the equations (\ref{ghzClass}) and of (\ref{extra}) are $\delta\in (0, \frac{\pi}{4}]$, $\alpha,\:\beta\:,\gamma\in (0, \frac{\pi}{2}]$ and $\varphi\in [0,2\pi)$. On the other hand, the standard form of $W$- class of sates is given as
\begin{eqnarray}
\label{wClass}
|\psi_{W} \rangle_{abc} = \sqrt{a}|001\rangle + \sqrt{b}|010\rangle + \sqrt{c}|100\rangle + \nonumber\\ \sqrt{d}|000\rangle,
\end{eqnarray}
where $a,b,c> 0$ and $d\equiv 1-(a+b+c)\geq 0$.
\\\\
The non-trivial tripartite entangled states (\ref{ghzClass}) can be converted  locally into the standard
$\vert GHZ\rangle_{abc}=\frac{\left(\vert 000\rangle_{abc} + 
\vert 111\rangle_{abc}\right)}{\sqrt{2}}$ state or vice versa by means of SLOCC, whereas the three parties
can locally transform the states (\ref{wClass}) into the state $\vert W \rangle_{abc}=\frac{
\left( \vert 001 \rangle_{abc} + \vert 010 \rangle_{abc} + \vert 100 \rangle_{abc} \right)}{\sqrt{3}}$ or vice versa under SLOCC \cite{durr622000}.
The in-equivalence between the states (\ref{ghzClass}) and (\ref{wClass}) means that if two arbitrary states $\vert \Phi\rangle$ and
$\vert \Psi\rangle$ can respectively be converted either into $GHZ$-class or $W$-class then it will not be
possible to transform $\vert \Phi\rangle$ into $\vert \Psi\rangle$ or vice versa under SLOCC. We already know that standard
$\vert GHZ\rangle_{abc}$ states are suitable for controlled dense coding while standard $\vert W\rangle_{abc}$ states
are not \cite{roy1550033}. Going further, we claim in this paper that if one can characterize a non-trivial tripartite state belonging to
$GHZ$-class (\ref{ghzClass}) then the state is useful for controlled dense coding under suitable choice of the state parameters,
whereas if the state belongs to the $W$-class of states, defined in eq.(\ref{wClass}), then they are not useful for CDC. This claim will have practical significance in  characterizing a
non-trivial tripartite entangled state belonging to any of these two in-equivalent class of states. The question is: given a state, how to confirm
that  it belongs to either of these two class ! Here the 3-tangle comes to rescue. The genuine tripartite entanglement
suggested by Coffman \textit{et. al} \cite{coffman612000} has the ability to distinguish between the above two in-equivalent
class of states. It was shown in \cite{durr622000,lohmayer2006,elt102008} that the 3-tangle (a measure of genuine
tripartite entanglement) is non-vanishing for any state in the $GHZ$-class (\ref{ghzClass}), while the 3-tangle vanishes for
any state in the $W$-class (\ref{wClass}).\\

\noindent For example, let us analyze the entanglement properties of the class of states $\vert \chi \rangle_{abc}$ of
eq.(\ref{xi1}), whose density matrix is represented by $\rho_{abc} = \vert \chi \rangle_{abc} \langle \chi \vert$,
the reduced density matrices are of the following form
\begin{eqnarray}
\nonumber
\rho_{a} = \left(\begin{array}{cc}
2\:k_{1}^{2}  & 0\\
0 & 2\:k_{1}^{2}\\
\end{array}
\right),~
\rho_{b} = \left(\begin{array}{cc}
k_{1}^{2}+\:k_{2}^{2}  & 0\\
0 & k_{1}^{2}+\:k_{2}^{2}\\
\end{array}
\right)
\end{eqnarray}
\begin{eqnarray}
\rho_{c} = \left(\begin{array}{cc}
 k_{1}^{2}+\:k_{2}^{2}  & 0\\
0 & k_{1}^{2}+\:k_{2}^{2}\\
\end{array}
\right)
\label{reduced}
\end{eqnarray}
where $k_{1}={\sin\theta}/{\sqrt{2}}$ and $k_{2}=\cos\theta/\sqrt{2}$ and consequently
$rank(\rho_{a}) = rank(\rho_{b}) = rank(\rho_{c}) = 2$. So the class of states may belong to
either of the two inequivalent classes ($GHZ$-class or $W$-class). It has been shown in the previous
sections that the state (\ref{xi1}) has nonvanishing 3-tangle and hence belongs to $GHZ$-class of states
(\ref{ghzClass}). The state (\ref{xi1}) is also useful for controlled dense coding as shown in Sec.$2$. Similar analysis can
reveal that the entangled states (\ref{xi2}), (\ref{xi3}) and (\ref{xi4}) belong to $GHZ$-class.
Thus we see that the states defined in equations (\ref{xi1}),  (\ref{xi2}), (\ref{xi3}) and (\ref{xi4})
actually belong to $GHZ$-class of states (\ref{ghzClass}) and hence are useful in controlled dense coding.

\section{Some other non-trivial tripartite entangled states and their utilities in controlled dense coding}\label{sec:example}

To support our conjecture that all non-trivial tripartite entangled states falling under $GHZ$-class with
non-vanishing tangle are actually useful in controlled dense coding for appropriate choices
of the state parameters, we cite some more examples below:
\subsection{Maximal sliced states in controlled dense coding}\label{subsec:example1}

Here we discuss another class of partially entangled states known as ``{\it maximal slice states}''
\cite{liumo2182,li052305} from the perspective of controlled dense coding. The three-qubit
partially entangled set of maximal slice (MS) states are defined as
\begin{eqnarray}
\vert MS\rangle_{abc} =\frac{1}{\sqrt{2}}(\vert 000\rangle_{abc} + \cos\alpha\vert 110\rangle_{abc}
+ \sin\alpha\vert 111\rangle_{abc}).
\label{mslice}
\end{eqnarray}
If $\rho^{MS}_{abc}$ denotes the density matrix corresponding to the state $\vert MS\rangle_{abc}$,
then it can be easily verified that $rank(\rho^{MS}_{a})=rank(\rho^{MS}_{b})=rank(\rho^{MS}_{c})=2$.
Also we see that the 3-tangle for MS state (\ref{mslice}) is found to be $\sin^{2}(\alpha)$ which is non-vanishing for $\alpha \neq n\:\pi$, where $n$ is any integer. Hence, we conclude that the states (\ref{mslice}) belong to $GHZ$-class of states (\ref{ghzClass}) and
are useful for controlled dense coding. Now, the effectiveness of such states (\ref{mslice}) in controlled dense
coding has already been investigated in \cite{liumo2182} which endorses our conjecture.
\subsection{Symmetric state in controlled dense coding}\label{subsec:example2}

Next, we consider three-qubit symmetric states \cite{you482297} defined as
\begin{eqnarray}
\vert S\rangle_{abc} = p \vert 000\rangle_{abc} +  q \vert 111\rangle_{abc} + r \vert 001\rangle_{abc} + s \vert 110\rangle_{abc},
\label{symmetry}
\end{eqnarray}
where $p^{2} + q^{2} + r^{2} + s^{2} = 1$. Without loss of generality, the coefficients $p$, $q$, $r$, $s$ are considered to be real
along with the condition $p \geq q \geq r \geq s$. Writing the three-qubit state (\ref{symmetry}) in density matrix form as $\rho^{S}_{abc}$,
one can verify that the rank of reduced density matrices are $rank(\rho^{S}_{a})=rank(\rho^{S}_{b})=rank(\rho^{S}_{c})=2$. Using
eq.(\ref{tangle}), we also find that the 3-tangle of the symmetric state is $4(p^{2}+r^{2})(q^{2}+s^{2})$ which is non-vanishing (since all of $p,\:q,\: r,\:s$ are not zero).
Hence, the states (\ref{symmetry}) must fall under the $GHZ$-class of states (\ref{ghzClass}). With the state (\ref{symmetry}), controlled
dense coding can be successfully performed as has been shown in \cite{you482297} which validates our claim.
\subsection{Non-prototypical $GHZ$- type - I and $GHZ$- type - II states in controlled dense coding}\label{subsec:example3}
Disclaimer: (\textit{The names $`$non- prototypical $GHZ$ states of type - I and II' used in this paper have been chosen just to distinguish between the following two forms of states. The identification of the states by these names are not universal}). 
\subsubsection{Type - I $GHZ$- class:}
The non-prototypical $GHZ$- type of state (named here as type -I $GHZ$- class), is of the following form
\begin{eqnarray}
\vert N_{I}\rangle_{abc} = L \{ \vert 000 \rangle_{abc} + l \vert 111 \rangle_{abc} \},
\label{pati}
\end{eqnarray}
with $L=\frac{1}{\sqrt{1+l^2}}$ and $l > 0$ is real. Such a state was proposed by Pati \textit{et. al} in \cite{patiparaagra}. Controlled dense coding was successfully shown with this state (obviously depending on the choice of the state parameter) in \cite{roy1550033}. Again using (\ref{tangle}), we see that the state (\ref{pati}) has tangle of the form $\tau=4 L^4 l^2$, which is obviously non-zero. It is clear from this expression that when $l=1$, the state has
tangle $\tau=1$. So, as expected, by our conjecture, this state is from $\vert GHZ\rangle$ class.

\subsubsection{Type - II $GHZ$- class:}

We have named a special class of states as type - II $GHZ$- class in this article. These states are basically transformed from $GHZ$- class of states. The type - II class of states are suitable for perfect teleportation and superdense coding and they can be converted from state $\vert \psi_{GHZ} \rangle_{abc}$ by a proper unitary operation. The unitary operation, however, is the tensor product of a two qubit unitary operation and a one qubit unitary operation as defined in \cite{liqiu2007}. We denote this transformed version of $GHZ$- states by $\vert N_{II}\rangle_{abc}$ and is defined as follows \cite{liqiu2007}
\begin{eqnarray}
\vert N_{II}\rangle_{abc}= \frac{1}{\sqrt{2}}[\vert \phi\rangle_{ab}\:\vert 0\rangle_{c}+e^{i\:\epsilon}\vert 00\rangle_{ab}\:\vert 1\rangle_{c}],
\label{Wnp1}
\end{eqnarray}
where
\begin{eqnarray}
\vert \phi\rangle = \frac{1}{\sqrt{n+1}}[\vert 10\rangle +\sqrt{n}e^{i\:\alpha}\vert 01\rangle]
\label{Wnp2}.
\end{eqnarray}
The state $\vert N_{II}\rangle_{abc}$ can be converted from $\vert \psi_{GHZ} \rangle_{abc}$ by 
\begin{eqnarray}
\vert N_{II}\rangle_{abc} = (U_{ab}\otimes I_{c})\vert \psi_{GHZ} \rangle_{abc}
\label{Wnp3}.
\end{eqnarray}
Here, $U_{ab}$ is a unitary operator acting on particles $a$ and $b$ given as
\begin{eqnarray}
U_{ab} &=& |\phi\rangle\langle 00| + |11\rangle\langle 01| + |\phi^{\perp}\rangle\langle 10| + e^{i\:\epsilon}|00\rangle\langle 11| \nonumber{}\\
|\phi^{\perp}\rangle &=& \frac{1}{\sqrt{n+1}}\:(\sqrt{n}e^{-i\:\alpha}|10\rangle - |01\rangle).
\label{Wnp4}
\end{eqnarray}
We have shown in our earlier work \cite{roy1550033} that, the states of the form (\ref{Wnp1}) are useful in controlled dense coding. Following the same rank analysis of the reduced systems and evaluating the non-vanishing 3-tangle of the
state (\ref{Wnp1}), one can confirm that they also belong to $GHZ$-class of states (\ref{ghzClass}). This in turn gives us one another example in justifying our conjecture.
\section{Conclusion}\label{sec:conclusion}

To conclude, we have shown that the class of non-trivial pure tripartite entangled states defined in eqs.
(\ref{xi1}),  (\ref{xi2}), (\ref{xi3}) and (\ref{xi4}) are useful for controlled dense coding which explores the
experimental significance of these states. We have shown that if a nontrivial class of three-qubit entangled states
fall into the category of $GHZ$-class of states defined in eq.(\ref{ghzClass}), then those states under
suitable choices of state parameters can be used in controlled dense coding. This operational criterion is mainly
determined by the genuine tripartite entanglement (3-tangle) of the state. To justify our claim further,
we have considered different types of three-qubit entangled states which are useful for controlled dense coding.
In future, one can generalize our result to other multi-partite systems in higher dimensions. It was shown that tripartite qutrit states of
$GHZ$-class are useful in controlled dense coding \cite{roy1550033}. So, investigating the operational criteria
for controlled dense coding in higher dimensional systems will also be interesting. To
generalize our present work for higher dimensions, one may consider {\it squashed entanglement} instead of
3-tangle \cite{sanders2012}.
\section*{Acknowledgements}
M. M. Ali acknowledges the support from the Ministry of Science and Technology of Taiwan and
the Physics Division of National Center for Theoretical Sciences, Taiwan.

\end{document}